\documentclass[twocolumn,aps,prb,amsmath,showpacs,tufte-book]{revtex4-1}
\usepackage{graphicx,amsmath,leftidx,amsfonts,subfig}
\usepackage{subfig,array,booktabs,diagbox}
\usepackage[normalem]{ulem}
\usepackage{dcolumn,bm,amssymb,indentfirst}
\usepackage{color,soul}
\setstcolor{green}
\usepackage{caption}
\begin{document}
\author{Bingyu Cui$^{1}$, Jonathan F. Gebbia$^{2}$, Michela Romanini$^{2}$, Svemir Rudi{\'c}$^{3}$, Ricardo Fernandez-Perea$^{4}$, F. Javier Bermejo$^{4}$, Josep-Lluis Tamarit$^{2}$,
Alessio Zaccone$^{1,5,6}$}
\affiliation{${}^1$Cavendish Laboratory, University of Cambridge, JJ Thomson
Avenue, CB3 0HE Cambridge,
U.K.}
\affiliation{${}^2$Grup de Caracterizacio de Materials, Departament de Fisica,
EEBE and \\ Barcelona Research Center in Multiscale Science and Engineering, \\
Universitat Politecnica de Catalunya, Eduard Maristany, 10-14, 08019 Barcelona,
Catalonia}
\affiliation{${}^3$ISIS Facility, Rutherford Appleton Laboratory, Chilton, Didcot, Oxfordshire OX11 0QX, United Kingdom}
\affiliation{${}^4$Instituto de Estructura de la Materia, C.S.I.C., Consejo Superior de Investigaciones Cient{\'i}ficas, Serrano 123, 28006 Madrid, Spain}
\affiliation{${}^5$Statistical Physics Group, Department of Chemical Engineering and Biotechnology, University of Cambridge, Philippa Fawcett Drive, CB3 0AS Cambridge, U.K.}
\affiliation{${}^6$Department of Physics ``A. Pontremoli", University of Milan, via Celoria 16, 20133 Milano, Italy}
\begin{abstract}

We applied the recently developed Generalized Langevin Equation (GLE) approach
for dielectric response of liquids and glasses to link the vibrational density of states (VDOS) to the dielectric response of
a model orientational glass (OG). The dielectric functions calculated based on the
GLE, with VDOS obtained in experiments and simulations as inputs, are compared with experimental data for the paradigmatic case of
2-adamantanone at various temperatures. The memory function is related to the
integral of the VDOS times a spectral coupling function $\gamma(\omega_p)$,
which tells the degree of dynamical coupling between molecular degrees of
freedom at different eigenfrequencies. With respect to previous empirical fittings, the GLE-based fitting reveals
a broader temperature range over which the secondary relaxation is active. Furthermore, the theoretical analysis provides a clear evidence
of secondary relaxation being localized within the THz ($0.5-1$ THz)  range of eigenfrequencies, and thus not too far from the low-energy modes involved in $\alpha$-relaxation. In the same THz region, the same material displays a crowding of low-energy optical modes that may be related to the secondary relaxation.

\end{abstract}

\pacs{}
\title{Secondary relaxation in the THz range in 2-adamantanone from theory and experiments}
\maketitle

\section{Introduction}

The dynamics of structural glasses (SG), those obtained by cooling or pressurizing the liquid state, is one of the major unsolved problems in condensed matter physics \cite{NGAI2000,debenedetti2001supercooled,Mauro19780,Roland_2005}. In addition to the inescapable collective structural ($\alpha$) relaxation, a faster ($\beta$) secondary relaxation often appears \cite{Stevenson2009,WTU2017,Vispa2017,Romanini2017Therm}. Such a process emerges in the susceptibility function as a separate peak or as an excess (shoulder) contribution to the main $\alpha$-relaxation. Johari and Goldstein \cite{Johari1971Viscous,JohariTJCP1970} revealed that such a process is an intrinsic dynamical process associated with non-cooperative local reorientations and must be differentiated from relaxations attributed to internal molecular degrees of freedom. In spite of the huge amount of experimental and simulations studies \cite{debenedetti2001supercooled,Roland_2005,Stevenson2009,WTU2017,Vispa2017,Romanini2017Therm,Johari1971Viscous,Ngai2001JPC,Caballero2017} as well as the existence of models \cite{DomschkePRE2011,Kwok1997_14a1,ngai1979_14a2,Ngai_2003_14a3,Gotze_1992} aimed at the understanding of this secondary relaxation, the physics behind is still under discussion. Two main interpretations are generally assumed, the existence of islands of mobility (involving only local regions, so a heterogeneous picture) and the alternative homogeneous interpretation in which molecules show small-angle reorientational diffusion. Whatever the interpretation assumed, the microscopic origin and the relation with low-energy eigenmodes of the system, i.e. soft optical modes, has not been investigated. This is commonly due to the difficulty to access different kinds of disorder appearing in SG, i.e., translational and orientational disorder, besides the conformational molecular disorder or internal molecular degrees of freedom. One strategy to simplify the problem is to reduce, as far as possible, the number of degrees of freedom of the studied system.

Under this premise, orientationally disordered (OD) phases (plastic crystal phases) giving rise to orientational glasses (OG) are not affected by the translational disorder \cite{Ramos1997,RBrand2002,Schneider2000,PuertasPRB2004} but still keep the orientational disorder large enough to be controlled. By further decreasing the disorder of the system, molecular systems displaying occupational well-defined disorder appear as those with less degrees of freedom \cite{Romanini2012,zuriaga2009new,PerezJCP2015,Romanini2016Double,TripathiTJCP2016,Hassine2016PCCP,gebbia2017glassy,Moratalla2019}. In these systems, molecules can occupy well-defined crystallographic sites for which the fractional occupancies are perfectly determined owing to the ergodic assumption. Moreover, for some cases \cite{Romanini2012,Negrier2014CGD}, a closely packed ordered phase exists for the same system and, thus, fundamental properties, such as those concerning thermodynamics or those related to the vibrational density of states (VDOS), are known and can be successfully compared.

This is the case of 2-adamantanone (C$_{10}$H$_{14}$O, hereinafter called $2$O$=$A). 2O=A is a "rigid" psedoglobular molecule of $C_{2v}$ symmetry obtained from adamantane by means of the substitution of two hydrogen atoms by one oxygen atom linked to a secondary carbon atom by a double bond.

The polymorphic behavior of this compound has been described many times in the literature in the past \cite{Romanini2012,Negrier2014CGD,SzewczykTJPC2015,RomaniniJPCM2017}. The OD room temperature phase melts at 529 K and exhibits a face-centered-cubic (FCC) structure (space group $Fm\bar{3}m$.) On cooling the OD phase, it transforms at around 178K to an "ordered" low-temperature monoclinic M phase (space group $P2_1/c$) with statistical disorder in the occupancy of oxygen atoms along three different sites (with fractional occupancies of 25\%, 25\% and 50\%), and returns to the OD phase on heating at ca. 205 K. In addition, submitting the 2O=A to a thermal cycling at normal pressure between 150 and 250 K (i.e. around the low-temperature to OD phase transition) a new denser low-temperature (stable) orthorhombic O phase (space group $Cmc2_1$) appears at the expense of the low-temperature (metastable) M phase and the transition temperature from the low-temperature O phase to the OD phase is found to be shifted from 205 K up to 221 K. Regardless of the occupational disorder of the M phase and the full ordered character of the O phase, specific heat capacities were determined to be strikingly close ~\cite{Negrier2014CGD}. This experimental evidence is coherent and comes from the close similarity of the experimental VDOS measured for both phases ~\cite{Negrier2014CGD}.  When comparing both VDOS, the fully ordered O phase shows sharper low frequency excitations than the occupationally disordered M phase at energies between 4 and 12 meV (i.e., 1 to 3 THz, the range comprised between low-energy phonons and the low energy localized frequencies, see Fig. 1). It should be noted that the low-energy range for those excitations was initially associated with some mixing of acoustic and optical modes. The excess excitations for the M phase with respect to the O phase then appear only in that low-energy range.

Despite the Kohlrausch stretched-exponential function and its Fourier transform analog Havriliak-Negami function provide still the most popular empirical functions to describe the slow $\alpha$ relaxation in any type of supercooled systems (SGs, OGs, etc), a different approach, which starts from first principles, was recently proposed \cite{CuiPRE2017}. This simple approach (see next section) was successfully applied to two orientational glasses, Freon 112 and Freon 113 \cite{CuiPRE2018}, for which the appearance of the secondary $\beta$ relaxation for Freon 112 was rationalized on the basis of lower and intermediate-energy excitations in the VDOS.

Here, we bring further insights into this complex problem by analyzing the paradigmatic case of 2O=A, i.e. a system with well-defined occupational disorder (M phase) and for which the fully ordered (O) phase is known. By using the modified theoretical model previously developed \cite{CuiPRE2018} we are able to account for the secondary relaxation appearing in the disordered M phase of 2O=A as well as the dynamical coupling of molecular disorder as a function of the eigenfrequencies, in particular, those concerning low-energy localized (optical) modes. Even more, we will demonstrate that the small differences in the VDOS between disordered M and ordered O phases are so subtle (just some optical modes shifted to lower energy for the disordered phase) that the dielectric susceptibility can be reproduced by using either of the two VDOS, despite the glassy features emerge due to the piling up of those optical modes (see also Ref.~\cite{Baggioli}) at low-energy in the occupationally disordered M phase.

\section{Inelastic neutron scattering measurements}
Vibrational features of monoclinic and orthorhombic phases of 2-adamantanone were studied by means of Inelastic Neutron Scattering (INS) experiments using the TOSCA spectrometer at the ISIS Pulsed Neutron and Muon Source of the Rutherford Appleton Laboratory (Oxfordshire, UK). In addition, DFT lattice dynamics calculations were performed for the stable fully-ordered orthorhombic phase to understand and interpret the INS data (see Supplemental Information~\cite{Supplementary} and Refs.~\cite{Vispa2017,Perdew,Payne,Pfrommer}).

The TOSCA indirect geometry time-of-flight spectrometer \cite{Parker_2014, PINNA201868} is characterized with high spectral resolution ($\Delta E/E \sim 1.25\%$) and broad spectral range ($-24 : 4000$ cm$^{-1}$). The sample was placed in thin walled and flat aluminum can (2 mm of thickness). The monoclinic phase was reached upon cooling from room temperature down below 178K and the stable orthorhombic one was found clycing between 150K and 240K for 6 times monitoring the evolution of the growth of the O phase over the M phase. The cycling was done with a controlled speed of heating/cooling (10K/min). For both phases, the sample chamber was cooled down to 10K by a closed cycle refrigerator (CCR) in order to reduce the effect of the Debye$-$Waller factor on the spectra, and the INS spectra were recorded. Finally, the data were converted to the dynamic structure factor, $S(Q,\omega)$, using the Mantid software framework.

\section{Theoretical GLE model}
Focusing on a tagged particle (e.g. a molecular subunit carrying a partial
charge which reorients under the electric field), it is possible to describe
its motion under the applied field using a particle-bath Hamiltonian of the
Caldeira-Leggett type, in the classical dynamics regime.
The particle's Hamiltonian is bi-linearly coupled to a bath of harmonic
oscillators which represent all other molecular degrees of freedom in the
system~\cite{Zwanzig1973}. Any complex system of oscillators can be reduced to
a set of independent oscillators by performing a suitable normal mode
decomposition. This allows us to identify the spectrum of eigenfrequencies of
the system, i.e. the experimental VDOS, with the spectrum of the  set of
harmonic oscillators forming the bath.

The particle-bath Hamiltonian under a uniform AC electric field, is given
by \cite{CuiPRE2017}: $H=H_P+H_B$ where $H_P=P^2/2m+V(Q)-q_e Q E_0\sin{(\omega
t)} $ is the Hamiltonian of the tagged particle with the
external electric field ($q_e$ is the charge carried by the particle),
$H_B=\frac{1}{2}\sum_{\alpha=1}^N\left[\frac{P_{\alpha}^2}{m_{\alpha}}+m_{\alpha}\omega_{\alpha}^2\left(X_{\alpha}
-\frac{F_{\alpha}(Q)}{\omega_{\alpha}^2}\right)^{2}\right]$ is the Hamiltonian
of the bath of harmonic oscillators that are coupled to the tagged
particle~\cite{Zwanzig1973}.
$H_B$ consists of two parts: The first part is the ordinary harmonic
oscillator; the second is the coupling term between the tagged particle
position $Q$ and the
bath oscillator position $X_{\alpha}$. The coupling function is taken to be
linear in the displacement of the particle,
$F_{\alpha}(Q)=c_{\alpha}Q$, where $c_{\alpha}$ is known as the strength of
coupling between the tagged atom and the $\alpha$-th bath oscillator.
Hence, there is a spectrum of coupling constants $c_{\alpha}$ by which each
particle interacts with all other molecular degrees of freedom in the
system. This spectrum of coupling strengths will play a major role in the
subsequent analysis.

As shown in previous work~\citep{CuiPRE2017,CuiPRE-FDT2018}, this particle-bath Hamiltonian
leads to a Generalized Langevin Equation (GLE) for the mass-rescaled coordinate
$q$ of the tagged particle:
\begin{equation}\label{eq:GLE}
\ddot{q}=-V'(q)-\int_{-\infty}^t \nu(t')\frac{dq}{dt'}dt'+q_eE_0\sin{(\omega
t)}.
\end{equation}
where the non-Markovian friction or memory kernel $\nu(t)$ is expressed in
terms of the spectrum of coupling constants $c_\alpha$ as
$\nu(t)=\sum_\alpha\frac{c_\alpha^2}{\omega_\alpha^2}\cos{(\omega_\alpha t)}$.

Then we can let the spectrum be continuous and $c_{\alpha}$ be a function of eigenfrequency
$\omega_p$ which leads to the following expression for the friction
kernel~\cite{Zwanzig1973}:
\begin{equation}\label{eq:nu}
\nu(t)=\int_0^{\infty}d\omega_p
D(\omega_p)\frac{\gamma(\omega_p)^2}{\omega_p^2}\cos{\omega_p t},
\end{equation}
where $\gamma(\omega_p)$ is the continuous spectrum of coupling constants, i.e.
the continuous version of the discrete set $\{c_{\alpha}\}$ and $D(\omega_p)\propto\sum_\alpha\delta(\omega_p-\omega_\alpha)$ is the continuous spectrum of vibrational frequencies, i.e. the VDOS.

The inverse
cosine transform in turn gives the spectrum of coupling constants
$\gamma(\omega_p)$ as a function of the memory kernel:
\begin{equation}\label{eq:gamma}
\gamma^2(\omega_p)=\frac{2\omega_p^2}{\pi
D(\omega_p)}\int_0^{\infty}\nu(t)\cos{(\omega_pt)}dt.
\end{equation}
This coupling function contains information on how strongly the single particle
motion is coupled to the motion of other particles in a mode with vibrational
eigenfrequency $\omega_p$. This is an important information, because it reveals
the degree of medium- and long-range (anharmonic) couplings in the motion of the
molecules.

Following  the same steps as those described in Ref.~\cite{CuiPRE2017}, we
obtain the complex dielectric function
\begin{equation}\label{eq:epsomega}
\epsilon^*(\omega)=1-A\int_{0}^{\omega_{D}}\frac{D(\omega_{p})}{\omega^2-i\omega\tilde{\nu}(\omega)-\omega_p^2}
d\omega_{p}
\end{equation}
where $A$ is an arbitrary positive rescaling constant, and $\omega_{D}$ is the
Debye cut-off frequency (i.e.
the highest eigenfrequency in the VDOS spectrum). As one can easily verify, if
$D(\omega_{p})$ were given by a Dirac delta, one would recover the standard
simple-exponential (Debye) relaxation.

\begin{figure}
\centering
\includegraphics[height=6cm ,width=8cm]{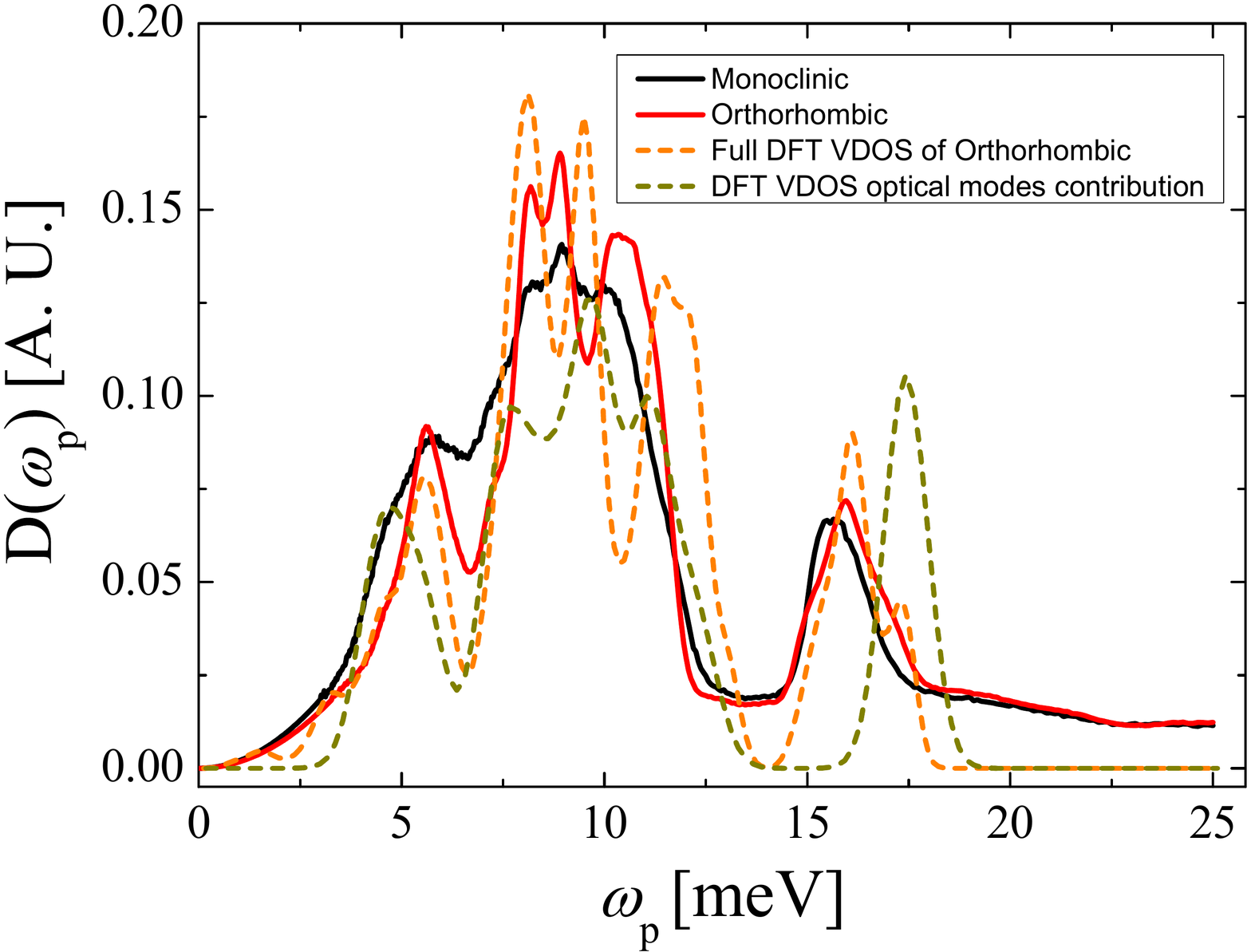}
\caption{Vibrational density of states (VDOS) from various approaches for 2-damantanone. Solid lines show experimental INS spectra, black for M phase and red for O phase. Dashed lines show VDOS from DFT lattice dynamic calculations of O phase, orange for VDOS using full phonon dispersion contribution and dark yellow for VDOS with only optical modes contribution.}
\label{fig:adaDOS}
\end{figure}

The VDOS is an important key input to the theoretical framework. The experimental VDOS, $D(\omega_{p})$, of M and O phases, measured at $T = 10K$ by means of inelastic neutron scattering using TOSCA spectrometer are
shown in Fig. \ref{fig:adaDOS}, together with VDOS obtained by DFT \textit{ab-initio} lattice-dynamics calculation for orthorombic phase at $\Gamma$-point (only optical modes contributes to VDOS) as well as using full phonon dispersions (see Supplemental Material~\cite{Supplementary}). This set of VDOS will be used as the input to explore the link between the vibrational spectrum and the dielectric response.

\begin{figure}[tb]
\centering
\includegraphics[height=5.5cm ,width=8cm]{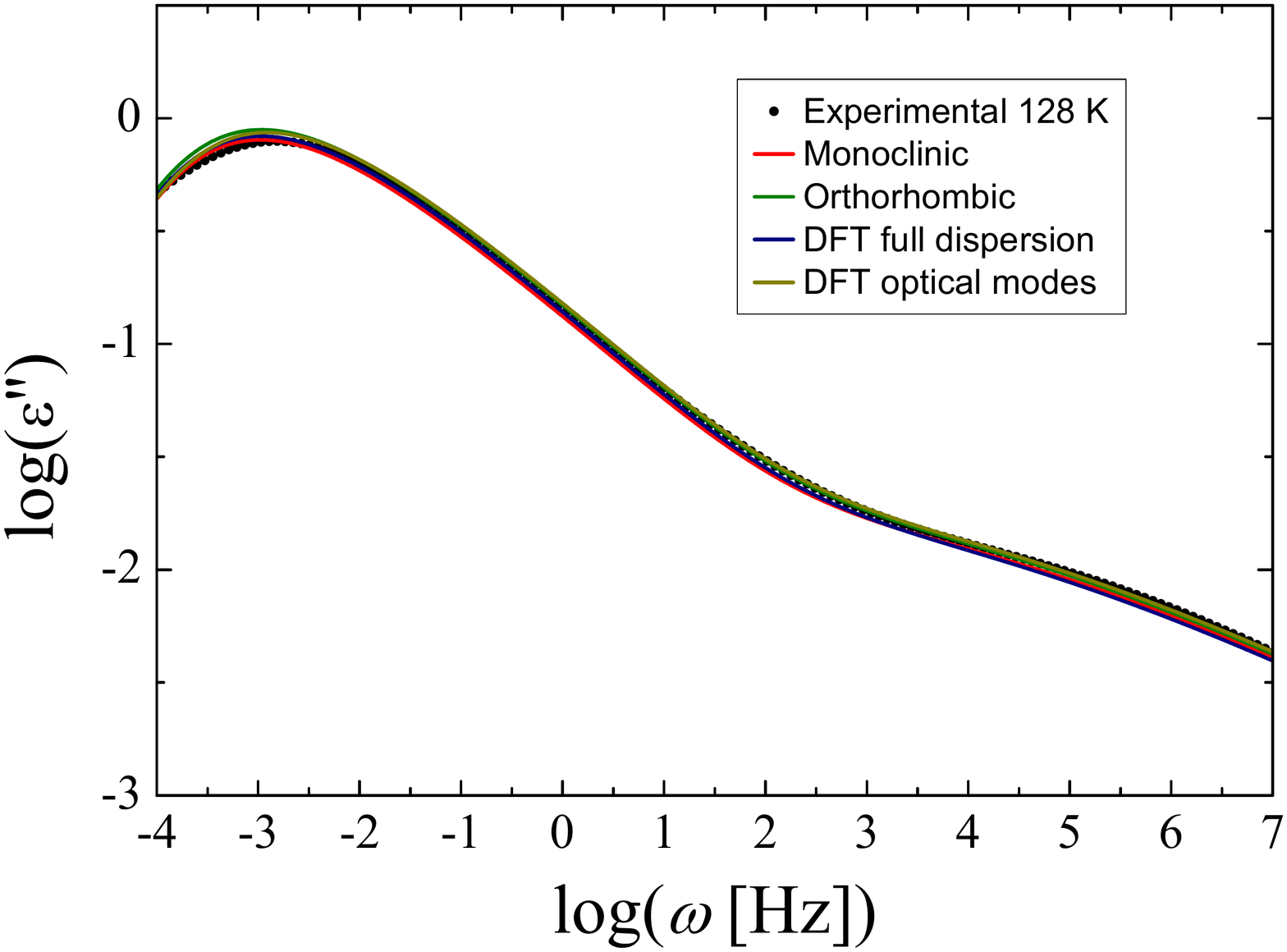}
\caption{Fitting of experimental data using the proposed theoretical model for
2-adamantanone at the same temperature with different VDOS as inputs.}
\label{fig:128K}
\end{figure}

\begin{figure}[tb]
\centering
\includegraphics[height=5.5cm ,width=8cm]{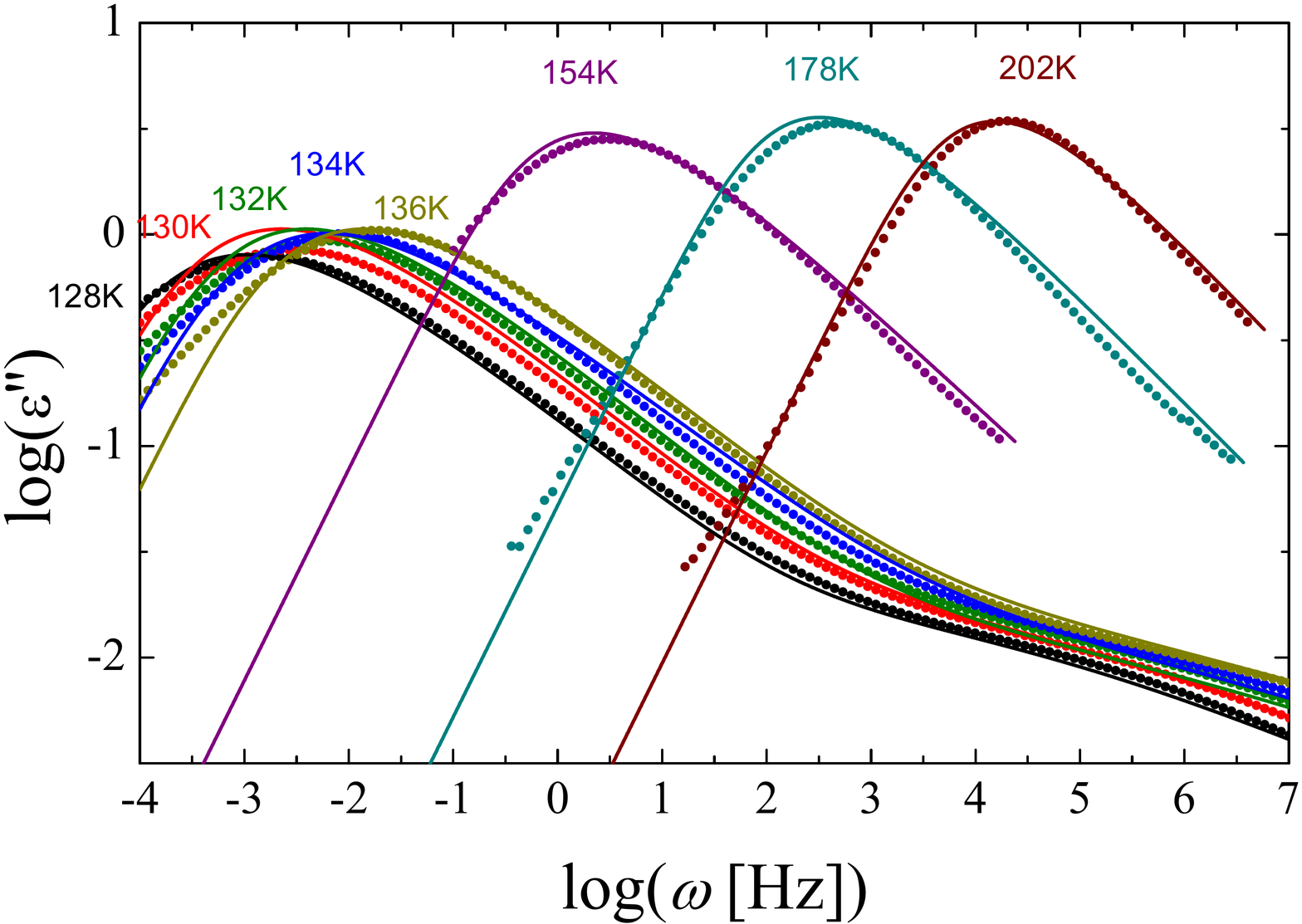}
\caption{Fitting of experimental data using the proposed theoretical model for
2-adamantanone at various temperatures with the same (experimental monoclinic) input VDOS.}
\label{fig:fitting}
\end{figure}

\begin{figure}[tb]
	\centering
	\includegraphics[height=5.5cm ,width=8cm]{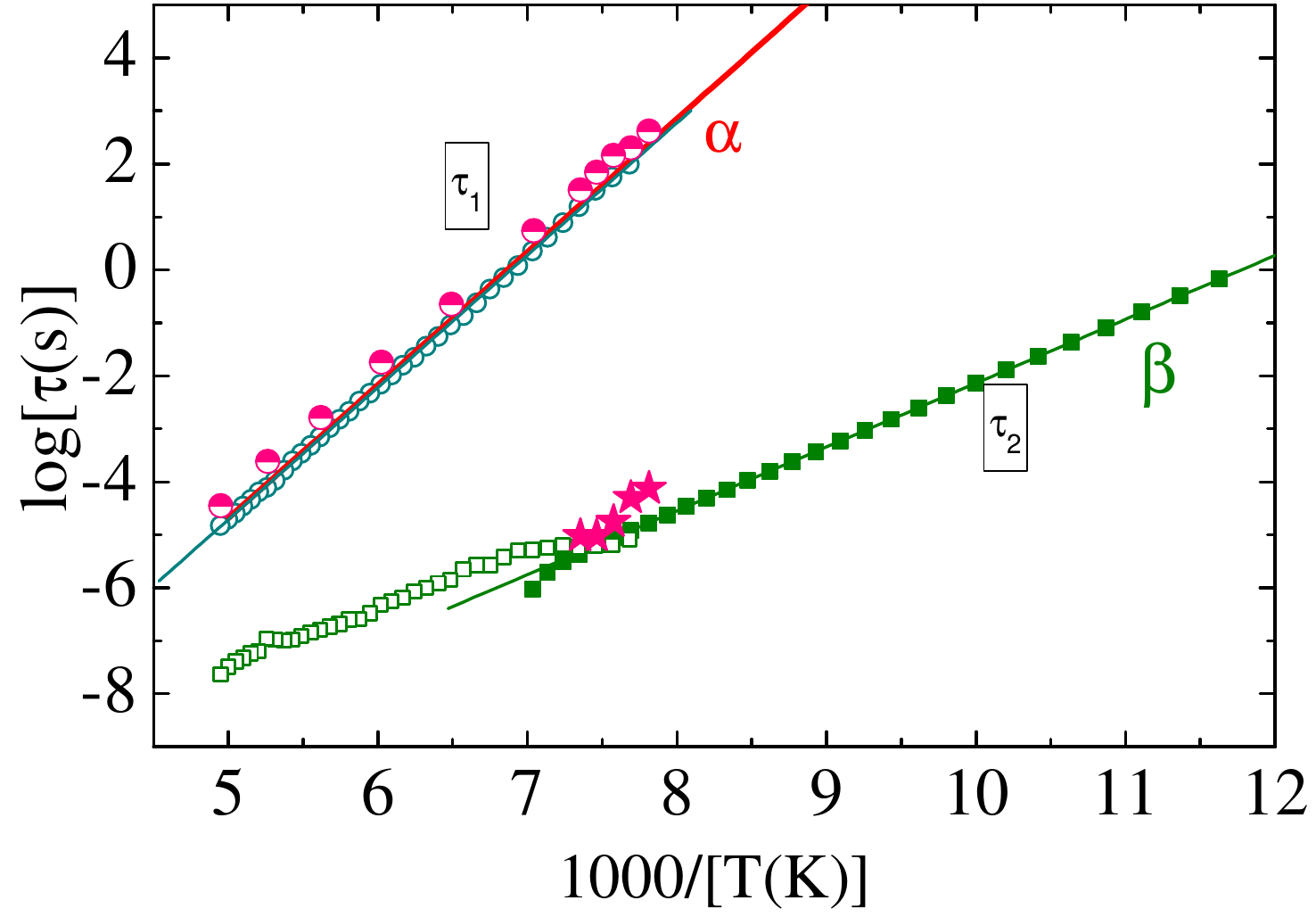}
	\caption{Relaxation times as a function of reciprocal of temperature for the different relaxation processes: Open circles: $\alpha$-relaxation (experimental values); open squares: $\beta$-relaxation (calculated values according to the CM model (Refs. \cite{ngai1979_14a2,Ngai_2003_14a3}); full green squares: $\beta$-relaxation (experimental values). Pink symbols are calculated according to our proposed theoretical model (see Table I): Full-empty circles for $\alpha$-relaxation and stars for $\beta$-relaxation.}
	\label{fig:LOGTAU}
\end{figure}

\begin{table*}[tbp]
\centering
\begin{tabular}{lcccccccc}
\hline
Temperature &128K &130K &132K &134K &136K &154K &178K &202K\\ \hline
$b_1$ &0.39 &0.40 &0.40 &0.38 &0.4 &0.46 &0.5 &0.53\\
$\tau_1$ (seconds)  &418.04 &200.47 &144.76 &69.09
&32 &0.222 &0.0016 &$3.50\cdot10^-5$\\
$b_2$ &0.225 &0.21 &0.19 &0.18 &0.18  & &  &  \\
$\tau_2$ (seconds) &$7.4\cdot10^{-5}$ &$5.00\cdot10^{-5}$ &$1.77\cdot10^{-5}$ &$9.60\cdot10^{-6}$ &$9.60\cdot10^{-6}$ & & \\
$\nu_2$ &0.023 &0.021 &0.02 &0.02 &0.025 &0 &0 &0
\\
\end{tabular}
\caption{\centering Parameters of the memory function.}
\end{table*}
\begin{figure}[tb]
\centering
\includegraphics[height=5.8cm ,width=8.8cm]{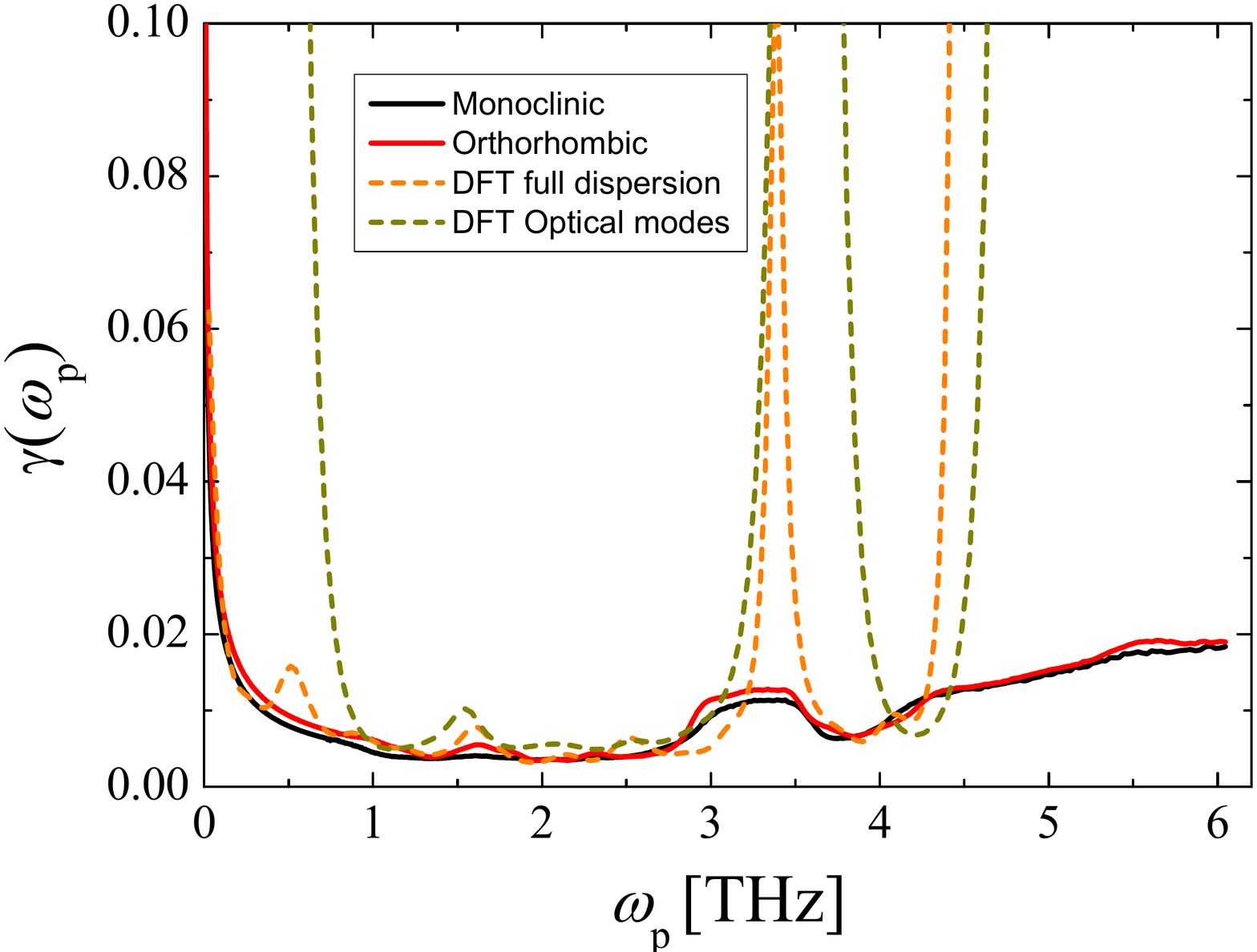}
\caption{
Spectrum of coupling constants as a function of the vibrational eigenfrequency
computed according to Eq. (\ref{eq:gamma}) using the same phenomenological memory
functions $\nu(t)$ used in the fitting of dielectric response in Fig.
\ref{fig:128K}, with the same color code used for the input VDOS of Fig \ref{fig:adaDOS} but the same form of memory kernel.}
\label{fig:gamma128K}
\end{figure}

\begin{figure}[tb]
\centering
\includegraphics[height=5.5cm ,width=8cm]{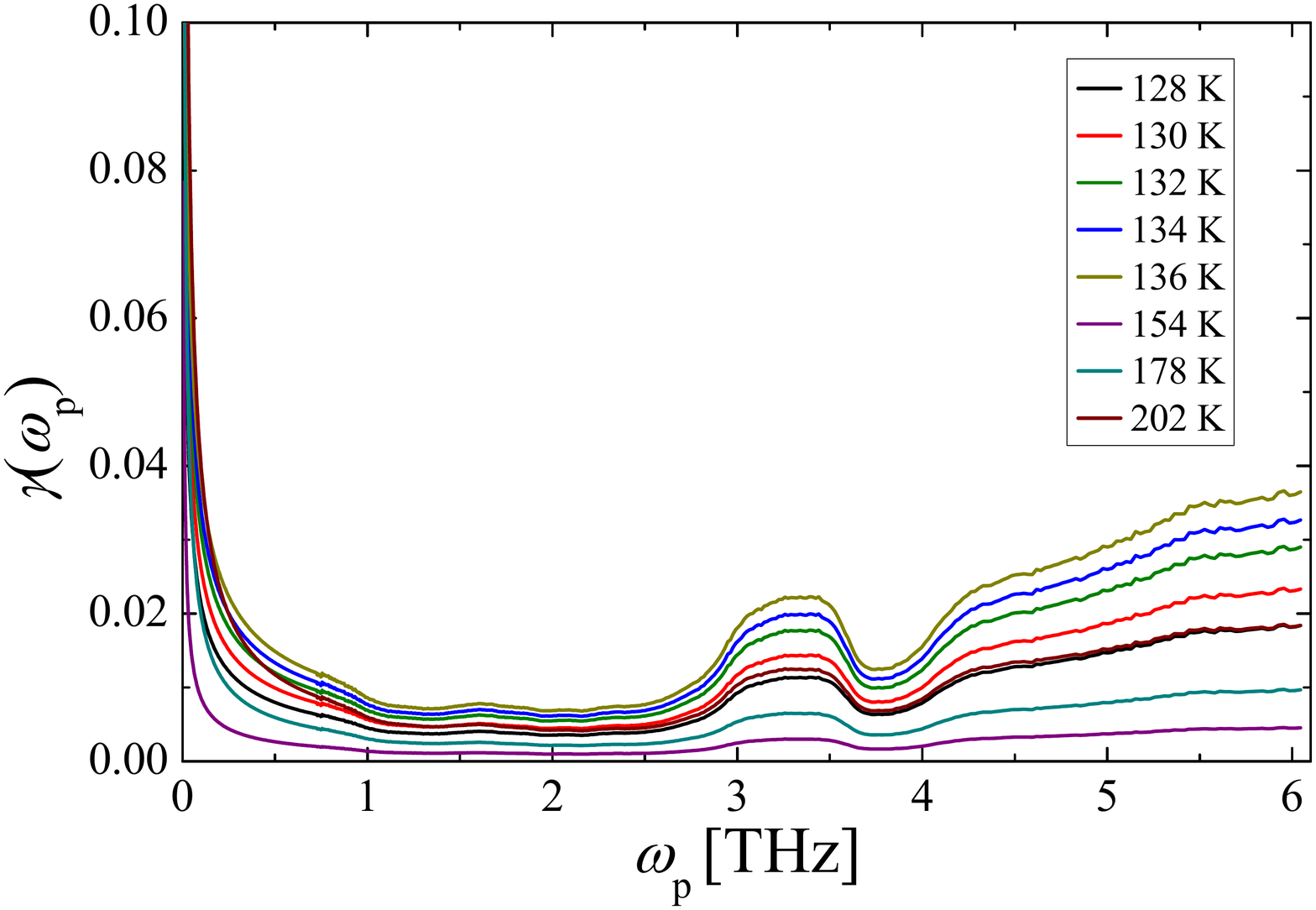}
\caption{
Spectrum of coupling constants as a function of the vibrational eigenfrequency
computed according to Eq. (\ref{eq:gamma}) using the phenomenological memory
functions $\nu(t)$ used in the fitting of dielectric response in Fig.
\ref{fig:fitting}, with same colour setting for the different temperatures.}
\label{fig:gamma}
\end{figure}

\begin{figure}
\centering
\subfloat[\centering Comparing the results of different weight $\nu_2$.]{
\begin{minipage}[b]{0.5\textwidth}
\includegraphics[width=0.8\textwidth]{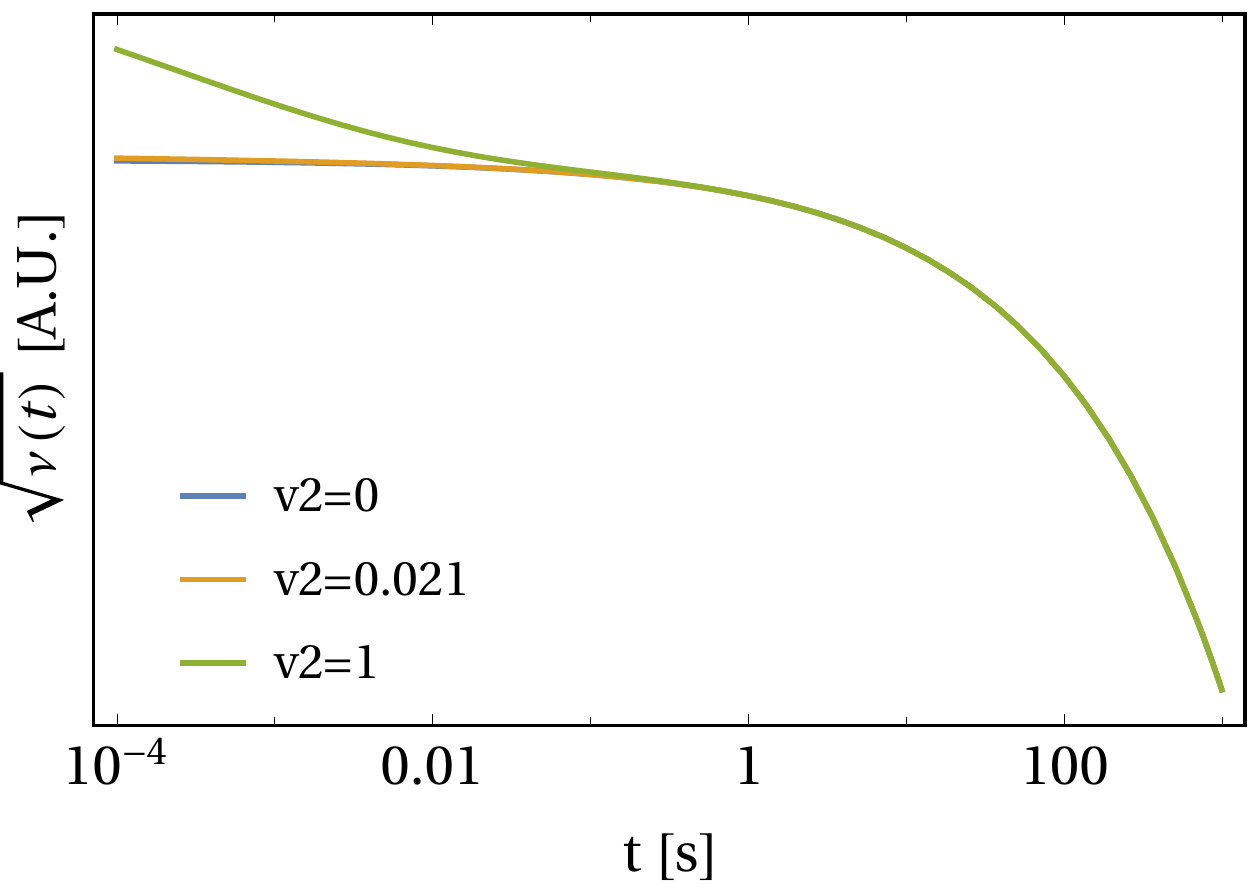}
\end{minipage}
}\\
\subfloat[\centering T=154K]{
\begin{minipage}[b]{0.5\textwidth}
\includegraphics[width=0.8\textwidth]{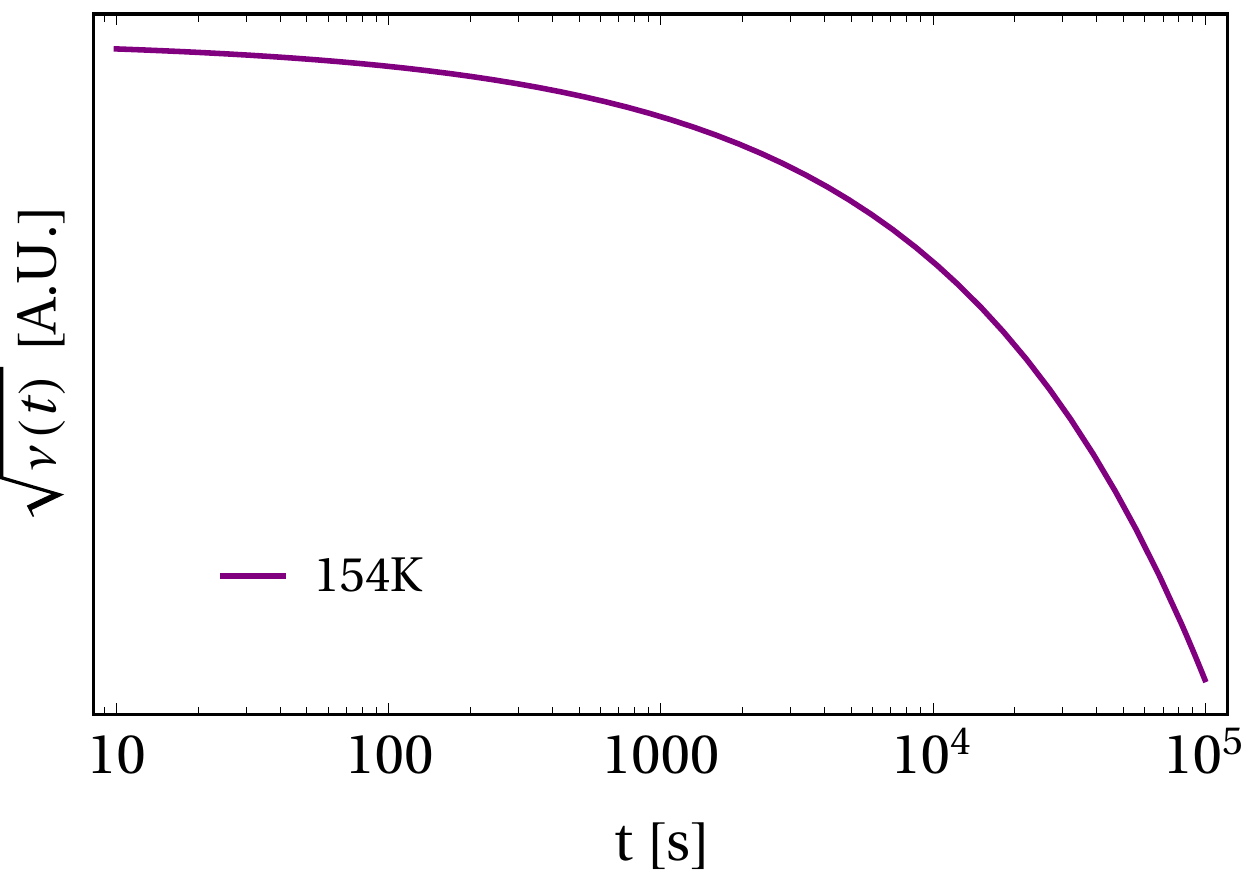}
\end{minipage}
}\\
\subfloat[\centering Higher temperature.]{
\begin{minipage}[b]{0.5\textwidth}
\includegraphics[width=0.8\textwidth]{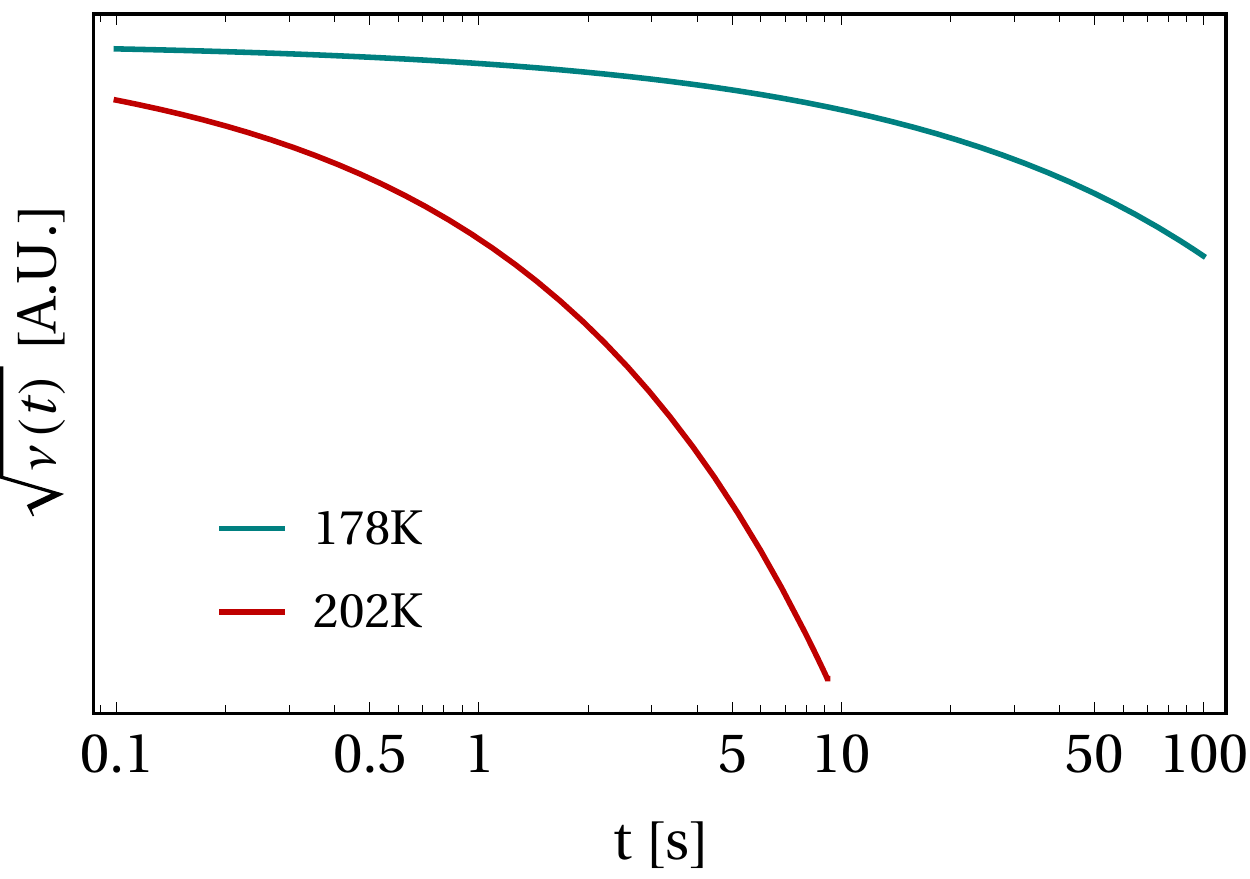}
\end{minipage}
}
\caption{\label{fig:ISF}Time decay of the square-root of total memory function
for the friction $\nu(t)$
according to the relation $F(q,t) \sim \sqrt{\nu(t)}$. Color settings are the
same as in Fig. \ref{fig:fitting}}
\end{figure}

\section{Results and discussion}
\subsection{Fitting of dielectric loss}
For data sets exhibiting secondary $\beta$-relaxation in loss modulus of dielectric
permittivity, we take the form of memory function $\nu(t)$ in Eq. (1) to be the sum of two terms,
both of which are stretched exponential. A motivation for the stretched-exponential form of the memory kernel comes from related approaches of Kia Ngai's coupling model~\cite{NGAI2000,Kwok1997_14a1,ngai1979_14a2,Ngai_2003_14a3}.
As in previous work for the case of Freon 112~\cite{CuiPRE2018}, we take the following
phenomenological expression for our memory function
\begin{equation}
\nu(t)=\nu_0\sum_i\nu_ie^{-(t/\tau_i)^{b_i}},
\end{equation}
where $\tau_i$ is a characteristic time-scale, with $i=1$ for $\alpha$
relaxation and $i=2$ for $\beta$ relaxation. $\nu_0$ is a constant prefactor while
$\nu_{i}$ with $i=1,2$ indicates the weight for the two different stretched exponentials. Without loss of generality, we set $\nu_1$ to be unity.
Fitting parameters at
different temperatures are listed in Table I.
To explore the effect of VDOS on dielectric relaxation behaviour, in Fig. \ref{fig:128K}, we only fit at one temperature at $T=128K$ with different VDOS data sets in Fig. 1, using the same memory kernel. The results are the same for other temperatures. The differences between the VDOS of the ordered and disordered phases are negligible, such that the procedure can hardly feel differences when comparing the dielectric loss (while keeping the memory kernel constant). The calculation using the VDOS calculated from DFT for the ordered phase also provides the same results. The result that the susceptibility is not so sensitive to the fine structure of VDOS might be the reason why the dielectric losses of some materials, such as ethanol \cite{Benkhof_1998}, levoglucosan\cite{Madejczyk2017} etc., in supercooled and plastic crystal phases show only subtle differences and in particular very small dfference for the $\beta$ relaxation. Conversely, if VDOS for some supercooled and plastic phases are similar, it means that orientational degrees of freedom dominate the system, i.e., the translational degrees of freedom (non-existent in the plastic phase) are not relevant, which has been demonstrated long time ago \cite{Ramos1997}.

Moreover, an important consideration is that concerning optical modes.
Fittings of the experimental data under various temperatures of the dielectric
loss modulus with the VDOS of the monoclinic phase are shown in Fig. \ref{fig:fitting}. For the fitting procedure, we have assumed that $D(\omega_p)$ and the overall
scaling in the stretched exponential, $\nu_0$, are temperature-independent. We used the algorithm in \cite{Wuttke2012} to perform the Fourier transform of stretched exponential functions. The so obtained relaxation times are plotted in Fig. 4.

\subsection{Analysis of spectrum of dynamic coupling constants}
To physically understand secondary relaxation in these systems, the spectrum of dynamical coupling
parameters (Eq. (\ref{eq:gamma})) has been analysed (see Fig.
\ref{fig:gamma128K} and \ref{fig:gamma}. In general, the coupling spectrum decays from the highest
Debye cut-off frequency of short-range high-frequency in-cage motions (above $5$THz), down to
the low eigenfrequency part where the coupling goes up with decreasing
$\omega_p$ towards zero, due to phonons or phonon-like excitations, which are
collective and long-wavelength and hence result in a larger value of coupling parameter $\gamma$.
The latter region of eigenfrequency is also the one corresponding to $\alpha$-relaxation. In the intermediate range of vibration frequency ($3-4$THz), fluctuations in the coupling spectrum are observed.

Looking at Fig. \ref{fig:gamma128K}, at about $0.5$THz, there is a clear peak for DFT full VDOS, which seems to become a "divergence" for the DFT optical modes due to the optical gap. Effectively, a maximum in VDOS corresponds to a minimum in the coupling and vice versa (whenever there is a gap in the VDOS there is a divergence in the coupling spectrum), which is clear in Eq. (\ref{eq:gamma}). Because of the sampling of the VDOS (with dispersions), at low energy there are some numerical fluctuations which cause a pronounced artificial peak at low frequency in coupling. In the case of the VDOS obtained without taking into account the dispersions of the branches (only optical modes contribute to VDOS), there is a gap from 0 to the first optical mode and this causes the divergence in the coupling spectrum. On the other hand, in the experimental orthorombic and monoclinic data, the coupling appears more attenuated and no divergences are observed.

\subsection{Analysis of secondary relaxation processes}
As shown in previous work~\cite{CuiPRE2018}, secondary relaxation shows up in the plot of dynamic coupling constant in an intermediate range comprised between low-energy phonons and the high-energy localized frequencies at the Debye cut-off.

Searching for the signature of the secondary relaxation process in Fig. \ref{fig:gamma}, we first note that the bump in the intermediate energy range 3-4THz cannot represent a genuine relaxation process because this bump is just the smeared-out version of a highly divergent feature visible in Fig. 5 and due to the presence of an optical gap in the VDOS in that energy range.

Hence, by exclusion, we can identify the emergence of the secondary relaxation as the other bump (or shoulder) in the plot just below $1$THz in Fig. \ref{fig:gamma}. A strong correlation of the $\beta$ relaxation with the amplitude of the nearly constant loss modulus related to the caged dynamics in the region of sub-THz has recently been shown \citep{Capaccioli2015}, whose relevance has been proven to explain why several high frequency quantities (including Mean Squared Displacement (MSD) from neutron scattering, or $\Gamma$ from Brillouin spectroscopy) related to caged dynamics amplitude show transitions at the temperatures where the $\beta$ relaxation and the $\alpha$ relaxation are frozen \cite{Ngai2015a, Ngai2015b,Wang2016}. The bump is not too far from the steeply increasing tail formed by collective modes active in the phonon-like regime and boson-peak soft modes responsible for $\alpha$-relaxation~\cite{CuiPRE2017}. This is also the reason why the contributions of $\alpha$ and secondary relaxation to the $\epsilon''$ curves are not always easily distinguishable, except for the lowest temperatures considered. In particular, it is interesting to observe the opposite behaviour of overall spectrum of coupling constant curves for temperatures above and below $T=136$K (which is comparable with glass transition temperature at $T_g\approx 132$K). The coupling spectra for temperatures below $T=136$K lie on top of the coupling spectra above
$T=136$K, which goes along with the fact that the data of dielectric loss in Fig. 3 are deprived of secondary relaxation processes above $T=136$K. Hence, the lack of secondary relaxation processes manifests as overall lower in magnitude values of dynamical coupling among degrees of freedom.

\subsection{Deducing the intermediate scattering function decays from dielectric loss fittings}
As shown by Sjoegren and Sjoelander~\cite{LSjogren1979} and as discussed in previous work~\cite{CuiPRB2018}, the time-dependence of
the memory function is proportional to the square of the time-dependence of the
intermediate scattering function (ISF), see also the discussion in ~Ref.~\cite{Baity}. Hence, for a process with pure $\alpha$
relaxation there is a simple stretched-exponential decay of the ISF, whereas
two decays are produced for systems with both $\alpha$ and $\beta$. To model
both $\alpha$ and $\beta$ relaxations under certain temperatures, we take
$\nu(t)$ to be the sum of two stretched exponential functions as in Eq. (5) above, with independent
parameters, which results in a two-step decay in the ISF.

This behaviour is shown in Fig.~\ref{fig:ISF} following the
relation $F(q,t) \sim \sqrt{\nu(t)}$. Since the weight $\nu_2$ used in the fitting for $\beta$-relaxation is small, we hardly see the
characteristic two-step stretched exponential decay of $F(q,t)$ present in systems with well
separated $\alpha$ and $\beta$ relaxations at low temperatures ($T\leq136$K). However, on the other hand, at high temperatures, the
$\alpha$ peak in $\epsilon''$, and the corresponding decay in $F(q,t)$, can be reduced to a
single characteristic time, as the time-scale range of the $\alpha$-relaxation
contains a strong
contribution from soft modes in the VDOS. This
is clear from Eq.~(\ref{eq:gamma}) where the term $\omega_p^{2}$ in the
denominator
gives a large weight to the low-$\omega_p$ part of the VDOS, which contains the
boson peak (excess over Debye $\sim \omega_{p}^{2}$ law) proliferation of soft modes.

\section{Conclusion}
In conclusion, we have  examined, for the paradigmatic case of 2-adamantanone, how its VDOS influences dielectric relaxation. We have found that: (i) only small differences in the VDOS between ordered and disordered phases exist, and can be seen only at very low energy; (ii) such differences concern mostly the optical modes, which for the monoclinic disordered phase are at lower energies; (iii) the piling up of the optical modes at low energy (at least in these organic compounds) are the root cause for the "glassy behavior", but in a very subtle way. Regarding the last point, the role of the optical modes is so subtle that a property directly linked to the VDOS such as the dielectric loss of the disordered phase (experimentally measured) and the "hypothetical" dielectric loss of the orthorhombic fully-ordered phase (not available experimentally) would be the same.
Finally, the fitting using the GLE theoretical model allows one to isolate the energy range of vibrational eigenmodes which participate in the secondary ($\beta$) relaxation. This is particularly important in systems like 2-adamantanone where there is no separation between $\alpha$ and secondary relaxations in the dielectric loss. It is seen that the vibrational eigenmodes responsible for secondary relaxation fall in an energy range around $0.5-1$ THz, which corresponds to the range where soft optical modes appear in the VDOS. This points at an unprecedented link between low-energy optical modes in organic molecular systems and secondary $\beta$ relaxation, which deserves further investigation in future work.

\begin{acknowledgements}
B. C acknowledges the financial support of CSC-Cambridge Scholarship. J.Ll. T.
acknowledges MINECO (FIS2017-82625-P) and the Catalan government (SGR2017-042)
for financial support.
\end{acknowledgements}

\bibliographystyle{apsrev4-1}
\bibliography{BK13564_revised}

\end{document}